\documentclass[journal]{new-aiaa}

\usepackage[utf8]{inputenc}

\usepackage{graphicx}
\usepackage{amsmath}
\DeclareMathOperator{\atantwo}{atan2}
\usepackage[version=4]{mhchem}
\usepackage{siunitx}
\usepackage{longtable,tabularx}
\usepackage{xcolor}
\usepackage{tikz}
\usepackage{siunitx}
\usepackage[ruled,vlined]{algorithm2e}

\title{Autonomous parafoil precision landing using convex real-time optimized guidance and control}

\author{Antoine Leeman\footnote{Young Graduate Trainee, GNC, AOCS \& Pointing Division. The author is currently affiliated with ETH Zürich. Email: aleeman@ethz.ch},Valentin Preda \footnote{Guidance, Navigation and Control Systems Engineer, GNC, AOCS \& Pointing Division},Irene Huertas\footnote{Guidance, Navigation and Control Systems Engineer, GNC, AOCS \& Pointing Division} ,Samir Bennani\footnote{Senior Advisor, GNC, AOCS \& Pointing Division}}
\affil{European Space Agency, Keplerlaan 1, 2201 AZ Noordwijk, Netherlands}

\newcommand{\tf}{{t_f}}
\newcommand{\tz}{{t_0}}
\newcommand{\tk}{{t_k}}
\newcommand{\tkp}{{t_{k+1}}}

\newcommand{\zz}{{z_0}}
\newcommand{\zf}{{z_f}}

\renewcommand{\t}{{^T}}
\renewcommand{\v}{\textbf}
\newcommand{\norm}[1]{\left\|#1\right\|_2 }
\newcommand{\dv}[1]{\dot{\v{#1}}}

\renewcommand{\i}{\int_\tz^\tf}

\newcommand{\ik}{\int_\tk^\tkp}
\newcommand{\norms}[1]{\norm{#1}^2}

\newcommand{\vh}{{{v}(z)}}
\newcommand{\vhc}{{v}}
\newcommand{\vvc}{{r}}

\newcommand{\vk}{{v(\tk)}}
\newcommand{\vkp}{{v(\tkp)}}
\newcommand{\vt}{{\tilde{v}_{k}}}

\newcommand{\ps}{\dot \psi_\text{max}}
\newcommand{\p}{\dot \psi(t)}

\newcommand{\D}{\Delta t_k}
\newcommand{\Wk}{W_k}

\newcommand{\da}{u(t)}
\newcommand{\pu}{\dot{\v{u}_r}(t)}

\newcommand{\vx}{\v x(t)}

\newcommand{\xk}{\v x_r(\tk)}
\newcommand{\xkp}{\v x(\tkp)}

\newcommand{\dx}{\dv x(t)}
\newcommand{\dpp}{\dv x_r(t)}

\newcommand{\xz}{\v x^0}
\newcommand{\pz}{\v x_r^0}

\newcommand{\pf}{\v x_r^f}

\newcommand{\vu}{\v u_r(t)}
\newcommand{\uk}{\v u_r(\tk)}
\newcommand{\ukp}{\v u_r(\tkp)}
\newcommand{\ul}{\bar{\v{u}}(t)}

\newcommand{\un}{ \bar{\v u}^{(n)}(t)}
\newcommand{\unp}{ \bar{\v u}^{(n+1)}(t)}
\newcommand{\unk}{ \bar{\v u}^{(n)}(\tk)}
\newcommand{\uso}{\v u_r^*(t)}
\newcommand{\usod}{\dot{\v u_r^*}(t)}
\newcommand{\duk}{\delta \ulk}

\newcommand{\uz}{\v u^0_r}
\newcommand{\uf}{\v u^f}

\newcommand{\lp}{\lambda_k^+(t)}
\newcommand{\lm}{\lambda_k^-(t)}

\begin{document}

\maketitle

\begin{abstract}
To overcome the limitations of current parafoil precision landing capabilities, an efficient real-time convex optimized guidance and control strategy is presented. 
Successive convexification of the parafoil guidance problem guarantees local optimality with polynomial convergence rate for efficient real-time implementation, where each iteration is dynamically feasible.
Our approach shows reliable and fast numerical convergence through in-flight recalculation of time of flight and a new optimal trajectory to cope with time-varying dynamics. 
The efficiency of our strategy is demonstrated via a comparative analysis of the existing X-38 in-flight demonstrated guidance and control system. Exhaustive Monte-Carlo simulations show performance improvements of about one order of magnitude. The concept proposed is simple, yet general, as it scales to any atmospheric parafoil landing system and allows efficient implementation relying only on the turn rate information.

\end{abstract}

\section*{Acronyms}

{\renewcommand\arraystretch{1.0}
\noindent\begin{longtable*}{@{}l @{\quad=\quad} l@{}}
SOCP & Second-Order Cone Programming\\
MPC & Model Predictive Control \\
WFF & Wind-Fixed Frame \\
SQP & Sequential Quadratic Programming \\
SCP & Sequential Convex Programming \\
IPM & Interior Point Method \\
SDP & Semi-Definite Programming\\
RRT & Rapidly-exploring Random Tree\\
DOF & Degree Of Freedom\\
LPV & Linear Parameter Varying\\
\end{longtable*}}
\section*{Nomenclature}

{\renewcommand\arraystretch{1.0}
\noindent\begin{longtable*}{@{}l @{\quad=\quad} l@{}}
$p_j$  & position along the j-axis \\
$\psi$ & heading angle of the parafoil \\
$z$ & altitude of the parafoil\\
$\vhc$ & horizontal speed\\
$\vvc$ & vertical speed\\
$v^.$ & horizontal ground speed\\
$w_j$ & wind along the j-axis \\
$\da$ & asymmetric deflection\\
$\rho$ & air density\\
$J$ & objective function for the trajectory optimization problem\\
$\boldsymbol \alpha$ & priority vector for the objective function\\
$\v x^f$ & targeted final state of the parafoil\\
$\v x^0$ & initial state of the parafoil\\
$\ps$ & maximum heading angle rate\\
$\tf$ & estimated final time of the trajectory\\
$u_j$ & j-component of the substituted input\\
$\lambda_k^{+/-}$ & interpolation function between the mesh point $k$ and $k+1$\\
$\D$ & time step between to mesh point $k$ and $k+1$\\
$A_d$ & state-transition matrix for the states\\
$B_k^{+/-}$ & state-transition matrices for the input\\
$\bar {\v u}$ & input trajectory used for the linearization

\end{longtable*}}

\section{Introduction}
\subsection{Background}
The increased availability of onboard computation power and efficient optimization algorithms associated with automated code generators\cite{Mattingley2012CVXGEN:Optimization} have dramatically boosted the industrial propagation of embedded optimized guidance and control strategies. Initiated by the research of Açikmeşe, numerous aerospace problems can be solved in real-time\cite{Mao2018AApplications}. With SpaceX and Blackmore, it was possible to land spaceships accurately 
\cite{Blackmore2017AutonomousRockets, Blackmore2010Minimum-landing-errorOptimization} using that strategy.
Convex optimization is widely used in real-time trajectory optimization problems thanks to the maturity of the technology \cite{Boyd2004ConvexOptimization,Ben-Tal2001LecturesOptimization} and the convergence speed to a global solution. The successive convexification (SCVx) method used in \cite{PadraigLysandrou2018ConvexConstraints} for instance, has been proven to be superlinearly convergent 
\cite{Mao2019SuccessiveProblems}.

The successes of the adoption of embedded onboard optimization relies on the numerical efficiencies of powerful algorithms and methods such as interior-point methods (IPM) for Second-Order Cone Programs (SOCP) and Semi-Definite Programming  (SDP). For further overview, the reader is referenced to [Yabar2020]. 

Originally, parafoil guidance has been performed using a sequence of open-loop trajectories relying on the concept of energy management. 
The various phases of flight are triggered by a trade-off between potential and kinetic energy indicators guiding sequentially the parafoil to its landing site. 
While this method is reliable and fast, it fails in taking into consideration the uncertainties of the parafoil and wind model, leading to poor landing precision performances. 
When the landing precision is a driving factor, especially in unknown and rugged terrain, such as for scientific payload delivery for exploration, or fairing recovery, the former technologies do not meet the requirements. To overcome these limitations, this paper demonstrates the potentials of using onboard optimized trajectory generation and control to increase drastically performance figures.

Driven by the need to embed the optimization onboard, series of authors have approached the problem from a different angle. To start with, the algorithm introduced  by Slegers in \cite{Slegers2009} uses a Model Predictive Control (MPC) on a 2-Degrees Of Freedom (DOF) model for the terminal guidance phase of the landing trajectory. The method can update the reference with an high refresh rate mitigating the effect of the highly unknown wind, which is assumed constant in the optimization process.

In \cite{Weinstein2018TrajectoryGuidance}, the authors use Particle Swarm Optimisation (PSO) to generate a trajectory online with an update rate of 1Hz. It can generate smooth trajectories thanks to the penalty on the heading acceleration, it shows good performances when compared against other real-time methods.

The idea exposed on \cite{Carter2009} embodies the actuator limitation in the optimization problem which allows both a refresh frequency of 1Hz and a guarantee of non-saturation of actuators.

The innovation of the guidance proposed in \cite{Handley2017EulerGuidance} is the use of Euler's elastica functions to get the trajectory that minimises the maximum curvature, taking advantage of the similarity of the equations of motion with a structure problem. Sacrificing a bit of precision, the algorithm run in real-time. The authors of \cite{Chiel2015HighGuidance} decide to tackle the wind uncertainty problem thanks to a new frame called the WFF (wind fixed frame).

The parafoil model used in
\cite{Luders2016WindParafoils} takes into account that the horizontal and vertical velocity are influenced by the altitude through the air density, leading to more representative results. The paper brings a method to take into account non-convex and difficult terrain while it ignores the winch dynamics.

In terms of collision avoidance,
\cite{LeFloch2017TrajectoryTerrain} presents an Rapidly-exploring Random Tree (RRT) algorithm to avoid the especially difficult urban environment.

Finally, \cite{Sun2020TrajectoryModel} uses a more realistic model compared to aforementionned methods. As a result, it takes into account the flared landing. However, the real-time implementation is not specified.\\

\subsection{Contribution}

Onboard optimization for parafoil guidance and control was considered by Slegers in \cite{Slegers2005}. While the linear MPC offers convergence and recursive feasibility guarantees, the low precision of the linear model used leads to a largely suboptimal solution, and lower landing precision performances.
On the other hand, the work of Rademacher in \cite{Rademacher2009} considers a more accurate model, including more physical parameters of the parafoil. It shows better results in terms of landing precision. Besides, it considers a realistic wind profile. However, the performances depends largely on the precision of estimation of the aerodynamic coefficients of the parafoil model. A small error on each parameter adds up to large landing precision error as shown in the Monte-Carlo simulations. Besides, the convergence guarantees are not clear for an arbitrary initial guess. The contribution of this paper is to provide a method with fast convergence guarantees with a 4-DOF model, which is intermediately accurate model with respect to the models used in the literature. In \cite{Mooij2021ConvexTitan} the authors use a sequential convex approach as well but relies on a 6-DOF model. While the resulting trajectory will be easier to track, each iterate is not guaranteed to be dynamically feasible. However, in our approach, since we use a 4-DOF model, the aerodynamic coefficients do not have to be known, and the method is fully transparent to parameters knowledge and estimation. The two main assumptions are the profile of air density and the level of saturation of the actuators. Overall, it enables the use of onboard optimization for a larger set of missions. It assumes the landing cone is free of any obstacle, ignoring any unevenness on the landing area. The landing cone is defined by the set of position where a dynamical feasible trajectory exist, leading to the landing site.
\subsection{Outline}
In the section \ref{sec:prob_formulation} the parafoil landing problem is introduced. It contains the parafoil and wind model used and the optimization problem to be solved and it provides an estimation of the speed and time of flight. Then, in the section \ref{sec:resolution}, a resolution of the aforementioned parafoil landing problem is presented. The resolution has the possibility to be implemented in real-time since each iterate consist of a convex problem and yields feasible solution to the original non-convex parafoil problem. The possibility to generate a new reference trajectory alleviates the effect of the uncertainties and unmodelled dynamics on the landing precision. Namely, the algorithm presented allows the continuous nonconvex problem to be discretized, convexified and linearized for real-time resolution. A simple controller tracking the generated reference trajectory is finally introduced. In the following section \ref{sec:results}, the performances are assessed with a Monte-Carlo simulation, as well as compared to the guidance of the X-38 mission.

\section{Problem formulation}
\label{sec:prob_formulation}
\subsection{Parafoil and environment modeling}
\begin{figure}[ht!]
    \centering
    \includegraphics{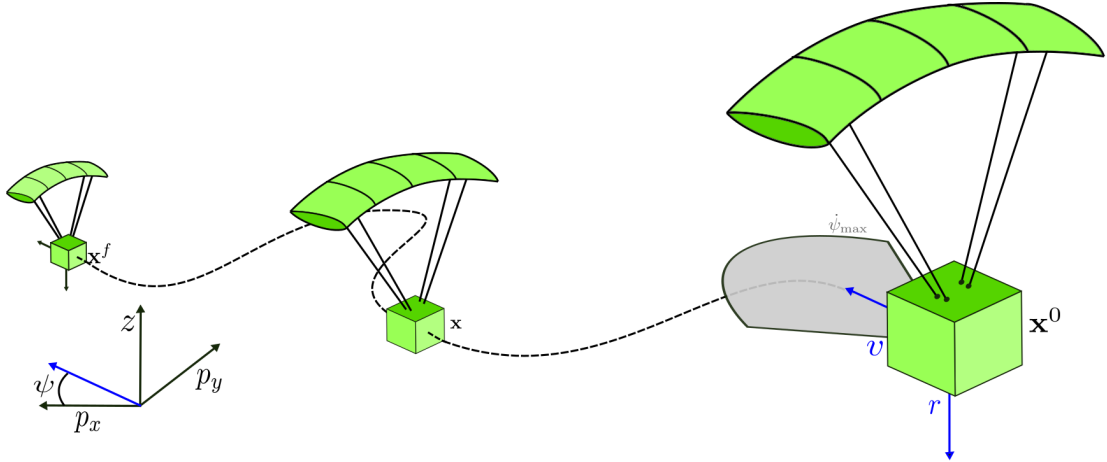}
    \caption{Schematic representation of the parafoil along the trajectory.}
    \label{fig:4dof_parafoil}
\end{figure}

\subsubsection{Parafoil}
Modeling the dynamical system, the parafoil, constitutes the first step of the problem formulation and there exists several ways to do it. First, we can use the first principles of physics and consider methodically all the forces and torques acting on the payload and parafoil, possibly considering highly nonlinear effects such as aeroelasticity. This classic modeling is accurate but contains many parameters and takes long to optimize over, which is redibitory for onboard optimization. It is commonly used as a benchmark model to verify the numerical performances of the guidance algorithm offline. Alternatively, we can base our model on the empirical observation that the parafoil rate of fall $\vvc$ and the horizontal speed $\vhc$ are roughly constant for a given air density, leading to a 4-DOF model. This kinematics model is illustrated on the Fig.\ref{fig:4dof_parafoil} and as in \cite{OlegYakimenkoNathanSlegers2015}, it depends on the wind strength $w$. This model is representative of the global parafoil motion and is therefore useful for guidance purposes. However, the 4-DOF model does not allow optimization for flare maneuver and generally ignores ground effect. Those dynamical limitations of the model will result in the sub-optimality of the solution of the parafoil landing problem. As it will be seen later, using a simpler model allows to run the optimization online. Both models, the one based on forces and torques (6-DOF) and and the overal motion (4-DOF) are compared in the Table \ref{tab:model_comp}.

\begin{table}[ht!]
\centering
\begin{tabular}{|r|l|l|ll}
\cline{1-3}
\multicolumn{1}{|c|}{\textbf{Model}} & \multicolumn{1}{c|}{\textbf{4-DOF}} & \multicolumn{1}{c|}{\textbf{6-DOF}} &  &  \\ \cline{1-3}
\cline{1-3}
\textit{Common use}                  & Trajectory optimization             & Simulation                          &  &  \\ \cline{1-3}
\textit{Limitations}                 & Roll and pitch angle                & Flexible modes                      &  &  \\ \cline{1-3}
\textit{Nature}                      & Kinematics                          & Dynamics                            &  &  \\ \cline{1-3}
\end{tabular}
\caption{Comparison between 4-DOF and 6-DOF parafoil model}
\label{tab:model_comp}
\end{table}

Choosing $p_j$ and $w_j$ as respectively the position and the wind speed in the j-direction, $\psi$ as the heading angle and $z$ as the altitude, the parafoil model used for the optimization is the following

\begin{equation}
\dx = f_{\text{4DOF}}(\vx) + B u(t) =
\begin{bmatrix}
\dot p_x(t)\\
\dot p_y(t)\\
\p\\
\dot z(t)\\
\end{bmatrix}
=
\begin{bmatrix}
&\vhc(z(t)) \cos \psi(t) + w_x(z(t))\\
&\vhc(z(t)) \sin \psi(t) + w_y(z(t))\\
&0\\
&\vvc(z(t))\\
\end{bmatrix}
+
\begin{bmatrix}
0\\0\\1\\0\\
\end{bmatrix}
u(t)
\label{eq:4dof}
\end{equation}
The control input $\da$ corresponds to an asymmetrical deflection of the parafoil trailing edges.
\subsubsection{Environment}
A precise model of the environment, such as the wind and the air density, will improve the representativeness of the problem and naturally improve the quality of the solution. In the optimization parafoil model of the Eq.(\ref{eq:4dof}), the horizontal and vertical velocities, depend on the air density modeled as a function of the altitude as
\begin{equation}
        \rho(z) = c_h(1-z\cdot c_\rho)^{c_e}
\end{equation}

where $c_h = \SI{1.225}{\kilo \gram \per  \meter \cubed }$, $c_\rho = \SI{2.256e-5}{\per \meter}$ and $c_e = \SI{4.2559}{}$
\cite{OlegYakimenkoNathanSlegers2015}.

As for the wind, it is generated thanks to a Dryden filter, similarly to\cite{Simplicio} as shown on Fig. \ref{wind_profile_pedro}. A Dryden filter is an Linear Parameter Varying (LPV) model $G_w$ which outputs the wind speed given the altitude and the ground speed as parameters. The input is a band-limited white-noise with unitary variance $n_w$. The wind generator is made up of 3 terms; a steady-state profile $G_{w_{ss}}$, a low-frequency/high-amplitude gust $G_{w_{LF}}$, a high-frequency/low-amplitude gust $G_{w_{HF}}$. There are parametrized with the altitude $z$ and the ground speed $v^.$. The resulting wind model used is
\begin{equation}
    G_w(s,z,v^.) = \frac{w_{x,y}(s,z,v^.)}{n_w(s)} = G_{w_{ss}}(z) + G_{w_{LF}}(s,z,v^.) + G_{w_{HF}}(s,z,v^.)
\end{equation}

To recover the wind in both horizontal directions, we use two distinct Dryden filters with different seed for the white noise\cite{Simplicio}. We consider the vertical wind speed to be negligible compared to the vertical speed of the parafoil. However, it could be easily integrated inside the time of flight estimation.
\begin{figure}[ht!]
    \centering
    \includegraphics{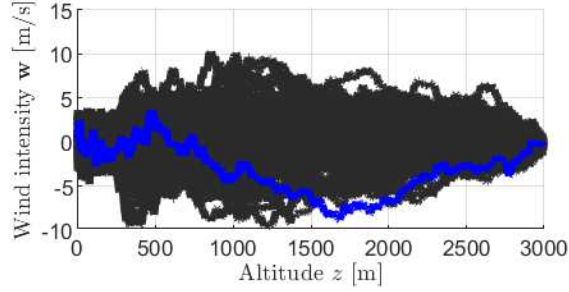}
    \caption{Set of wind profiles generated by the Dryden filer (black) where a single profile is outlined (blue).}
    \label{fig:wind_profile_pedro}
\end{figure}

\subsection{Optimization problem}
The parafoil landing problem consists in finding an optimal time-varying reference control input $\da$ for the asymmetrical deflection, leading the parafoil towards the landing site with a desired heading angle, minimizing the quadratic control cost for higher control margins. The desired final state $\v x^f$ depends on the mission. The objective of the problem is captured by the cost function (\ref{eq:cost_fun}) that we seek to minimize. Formally, the solution of the problem are the functions of time $\v x^*(t)$ and $u^*(t)$ which are required to respect a series of constraints\footnote{The $.^*$ is used to disambiguate the solution of the problem from its variables.}. We will see in the following that $u^*(t)$ is used as a feed-forward, or guidance, while $\v x^*(t)$ is used for the feedback control.
As such, the constraint (\ref{eq:dynopt}) enforces the states $\vx$ and control variable $\da$ to respect the 4-DOF dynamics of the parafoil, leading the solution to be dynamically feasible. The constraint (\ref{eq:init_cond}) shows that the initial states value are imposed at the initial time $\tz$ where $\v x^0$ stems from the navigation's sensors. The actuation faces a physical saturation and has to be bounded between $-\ps$ and $\ps$ as depicted by the constraint (\ref{eq:prob}). An accurate estimation of the final time $\tf$ is derived in the section \ref{sec:Altitudeevolution}, decorrelating the horizontal and vertical channel of the problem. The parafoil landing problem writes
\begin{subequations}
\begin{align}
\underset{\vx, \da}{\text{minimize }}&J  =  l_f(x(\tf))  + \i \da^2 dt         &\textit{\quad cost function} \label{eq:cost_fun}\\
\text{subject to }&  \dx =  f_{\text{4DOF}}(\vx) + B u(t)                                  &\textit{\quad dynamics}\label{eq:dynopt}\\
                  &  \v x(\tz) =  \v x^0                                               &\textit{\quad initial conditions}\label{eq:init_cond}\\
                   & -\ps \le \da \le \ps                                                               &\textit{\quad control input saturation}
\label{eq:prob}
\end{align}
\label{eq:opt_problem}
\end{subequations}
where the parameters $\tf$, $v$ and $r$ are derived in the following section (\ref{sec:param}). We define

\begin{equation}
     l_f(x(\tf)) = \alpha_1 ||[p_x(\tf), p_y(\tf)]\t -  x^f||_2 + \alpha_2 ||\psi(\tf) - \psi^f ||_2
\end{equation}

As we seek to minimize both the control energy and the missed final state $\v x^f$, the parafoil landing problem is multi-objective. The vector $\boldsymbol\alpha = \left[\alpha_1\quad \alpha_2\right]^T$ is a numerical artifact to select a single point of the pareto front of the multi-objective optimization problem.
\begin{equation}
\alpha_1 >> \alpha_2>>1 
\end{equation}
This way, this $\boldsymbol\alpha$ selects a trajectory that reaches $\v x^f$ at the lowest control cost. If the dynamics of the parafoil does not allow it to reach $\v x^f$, it will minimize the missed distance $\v x(\tf) - \v x^f$, at whatever control cost. It is worth highlighting that the optimization problem will always have a dynamically feasible solution, since the final state appears only in the objective function. However, it can be difficult to obtain because of the non-convexity of the constraint (\ref{eq:dynopt}), preventing the use of a convex solver. The procedure to convexify those constraints is outlined in the section \ref{inputsubsitution}. Finally, the optimization problem is continuous and cannot be understood by a numerical solver. The discretization procedure is explained in the section \ref{sec:discretization}.

\subsection{Parameters}

\label{sec:param}
\subsubsection{Speed}
The 2 most important parameters of the model \eqref{eq:4dof} are the horizontal $v$ and vertical speed $r$. Because the density of the air decreases with the altitude, both $v$ and $r$ will be attitude dependent and considering this variation will improve the fidelity of the model and subsequently the landing precision. To determine their functions of the altitude, respectively $v(z)$ and $r(z)$, we assume the lift force constant and equal to the weight along the trajectory and the quantity $\rho v^{.2}$ will remain unchanged\cite{LeFloch2017TrajectoryTerrain} where $v^. = \sqrt{r^2 + v^2}$. Based on the measure of the speeds and density at the initial altitude $\zz$, respectively, $\vhc(\zz)$, $\vvc(\zz)$ and $\rho(\zz)$, we have the value of the horizontal and vertical speed at any altitude $z$
\begin{equation}
    \vhc(z) = \vhc(\zz)\sqrt{\frac{\rho(\zz)}{\rho(z)}}\quad  \text{and}\quad   \vvc(z) = \vvc(\zz)\sqrt{\frac{\rho(\zz)}{\rho(z)}}. 
\end{equation}

\subsubsection{Altitude evolution}
\label{sec:Altitudeevolution}

Because the vertical speed $r$ is now known at any altitude, there is a direct relation between the altitude and the time elapsed. Thanks to this relation, we will be able to transform the control $\da$ as a function of the altitude $u(z)$. This is an asset since the wind is considered known as a function of the altitude and it will contribute heavily to the dynamics of the system. We can define the following change of variables

\begin{equation}
\begin{aligned}
    t(z) &= \tz + \int_\zz^z \vvc(\eta) d\eta = \tz+ \frac{ \int_\zz^z \sqrt{\rho(\eta)} d\eta}{\vvc(\zz) \sqrt{\rho(\zz)}}\\
        &= \tz+ \frac{\sqrt{c_\rho}}{\vvc(\zz) \sqrt{\rho(\zz)}}\left[ \frac{(1-\zz c_z)^{c_e/2+1}}{(c_e/2+1)c_h} - \frac{(1-z c_z)^{c_e/2+1}}{(c_e/2+1)c_z}\right]\\
\end{aligned}
\end{equation}
The relation can also be inverted to recover the corresponding altitude as a function of time. The idea to exchange the time and altitude variable is also used in \cite{Weinstein2018TrajectoryGuidance}.
\begin{equation}
    \begin{aligned}
    z(t)&= \frac{1}{c_z}\left( 1- \sqrt[c_f]{\left[\frac{(1-\zz c_z)^{c_f}}{c_fc_h} - (t-\tz) \left(\frac{\sqrt{c_\rho}}{\vvc(\zz) \sqrt{\rho(\zz)}} \right)^{-1}\right]c_fc_z }\right)\\
\end{aligned}
\end{equation}

Moreover, we can evaluate this function for the final altitude $\zf$ to get an estimation of the fixed final time of the trajectory.
\begin{equation}
            \begin{aligned}
     \tf = t(\zf) &= \tz+ \frac{\sqrt{c_\rho}}{\vvc(\zz) \sqrt{\rho(\zz)}}\left[ \frac{(1-\zz c_z)^{c_e/2+1}}{(c_e/2+1)c_h} - \frac{(1-\zf c_z)^{c_e/2+1}}{(c_e/2+1)c_z}\right]\\
\end{aligned}
\end{equation}

The final time depends only on the current air density and vertical speed, which can both be measured in-flight. Having a variable final time would constitute another non-convexity to get rid of, as shown in \cite{Szmuk2018SuccessiveFree-Final-Time}.

\section{Resolution}
\label{sec:resolution}
\subsection{Approach}

The parafoil landing problem as introduced in the Eq.(\ref{eq:opt_problem})is non-convex because of the trigonometric functions. The general idea of the resolution will be to substitute the non-convex functions by constrained linear functions. As such, to exploit the convergence speed of the resolution of convex problems, we will solve the problem via a Sequential Convex Programming (SCP) algorithm similarly as in \cite{Reynolds2020AGuidance}. For every iteration, we solve a convexified version of the problem, linearized around the previous solution. We initialize with a dynamically feasible zero-input trajectory and the problem is then sequentially solved until convergence. At each iteration, the solution provided by the algorithm is dynamically feasible and this algorithm is guaranteed to converge to a numerical feasible solution\cite{Mao2019SuccessiveProblems}. Because each iteration is dynamically feasible and each iteration is numerically feasible, the output of the algorithm is always a dynamically feasible trajectory. Besides, since each iteration of the algorithm is dynamically feasible, we can choose to stop the convergence process at any given time and output a suboptimal solution to comply with the real-time constraints. Each iteration is a convex problem, which can be solved in polynomial time. The relative value of the cost function between two consecutive iterations is chosen as a stopping criteria for the iterative algorithm. However, the parafoil landing problem is continuous and its solution has an infinite dimension. To be computationally tractable, it is finally shown how to discretize the problem by constraining the input function $u(t)$ to be piece-wise linear.

\subsubsection{Input substitution}
\label{inputsubsitution}

To form the convexified problem, we introduce a change of variables and a quadratic constraint to remove the inherently non-convex trigonometric functions. A similar method was used in \cite{Chiel2015HighGuidance}.

\begin{equation}
    \begin{aligned}
    \vu := \begin{bmatrix}
    u_x(t)\\
    u_y(t)\\
    \end{bmatrix}
    = 
    \begin{bmatrix}
    \vhc(z(t)) \cos \psi(t)\\
    \vhc(z(t)) \sin \psi(t)\\
    \end{bmatrix}
    \end{aligned}
    \quad\text{and}\quad
    \norm{\vu} = \vhc(z(t))
    \label{eq:nonL_cons}
\end{equation}
Likewise, we define a new state vector $\v x_r(t) = \left[ p_x(t) \quad p_y(t)\right]^T$ and the heading angle $\psi$ disappeared from the initial state vector. After a few algebraic operations, the parafoil landing problem (\ref{eq:opt_problem}) is equivalent to

\begin{subequations}
    \begin{align}
\underset{\v x_r(t), \vu}{\text{ minimize }} J &=\boldsymbol\alpha\t 
    \begin{bmatrix}\label{eq:obj}
    \norm{\v x_r(\tf) - \pf}\\
    \v u(\tf)\t \uf+1\\
    \end{bmatrix} + \i \arctan^2\left(\frac{\vu \times \pu }{\vu\t \vu}\right) dt & \textit{cost function}\\
\text{subject to }        
&\dpp = \vu + \v w(z) \label{eq:dyn} &\textit{dynamics}\\
&    \norm{\vu} - \vhc(z(t)) = 0 \label{eq:vh}&\textit{substitute input constraint}\\
& \arctan^2\left( \frac{\norm \pu}{\norm \vu}\right) \le \ps^2 \label{eq:u_max}&\textit{input saturation}\\ 
&\v x_r(\tz) = \pz&\textit{initial conditions - position}\\
&    \v u_r(\tz) = \uz& \textit{initial condition - heading angle}
\end{align}
\label{eq:cvx_pa}
\end{subequations}

where the parameters $\uz$, $\pz$, $\pf$ and $\uf$ are inferred from the problem (\ref{eq:opt_problem}).

The solution of the Problem \ref{eq:opt_problem} $\{ u^*(t), \v x^*(t)\}$ can be recovered from the solution of the Problem (\ref{eq:cvx_pa}) $\{\v u_r^*(t), \v x_r^*(t)\}$ as

\begin{equation}
    u^*(t)= \frac{d}{dt}\atantwo\left(\frac{\uso \times \usod }{\uso\t \uso}\right)
    \quad
    \v x^*(t)=
    \begin{bmatrix}
    \v x_r^*(t)\\
    \atantwo\left(\frac{\uso \times \usod }{\uso\t \uso}\right)\\
    r(z(t))\\
    \end{bmatrix}
    \label{eq:conv_states}
\end{equation}
The estimation of the wind profile $\v w(z)$ goes beyond the scope of this article. As emphasized in \cite{Chiel2015HighGuidance}, it is important to consider the wind proactively rather than reactively to achieve better landing precision performances. 
\subsubsection{Discretization}
\label{sec:discretization}
Standard numerical tools such as \cite{Diamond2016CVXPY:Optimization} require the variables of the optimization problem to be real numbers. Indeed, searching on the set of all possible functions would be intractable. To convert the continuous problem into a discrete counterpart, we define a temporal mesh and impose the constraints only on its nodes. Motivated by \cite{Malyuta2019DiscretizationProblem}, we choose to constrain the input function to the following format
\begin{equation}
    \vu = \lm \uk + \lp \ukp \quad \forall t \in [\tk, \tkp)
    \label{eq:input_piecewise_linear}
\end{equation}

which interpolates linearly the value between $\uk$ and $\ukp$ with

\begin{equation}
    \lm = \frac{\tkp - t}{ \D} \quad \lp = \frac{t - \tk}{\D}
\end{equation}
Although $\D$ was selected constant for the simulations, we could select a more dense temporal mesh closer to the landing point. The drastic reduction of the dimension of the problem introduced by this discretization results in an approximated solution. Given the nature of the problem, its solution is smooth and well fitted by a piece-wise linear function provided that the mesh is dense enough. In practice, increasing iteratively the number of nodes, the quality of the solution did not improve further beyond 40 nodes, indicating the convergence to the continuous problem's solution for a fixed final time.

In the following, for the sake of simplicity, the subscript will refer to the corresponding time value, such as, $ \v x_{k} = \xk$ and $ \v u_k = \uk$. Using this input format, we can derive an approximation of the objective as a quadratic function from Eq.(\ref{eq:obj}).
\renewcommand{\vk}{{v_k}}
\renewcommand{\vkp}{{v_{k+1}}}
\renewcommand{\uk}{\v u_k}
\renewcommand{\ukp}{\v u_{k+1}}
\renewcommand{\xkp}{\v x_{k+1}}
\renewcommand{\xk}{\v x_{k}}
\renewcommand{\unk}{\bar{\v u}^{(n)}_k}
\renewcommand{\duk}{\delta \bar{\v u}^{(n)}_k} 
\begin{equation}
        J_2 =  \i \arctan^2\left(\frac{\vu \times \pu }{\vu\t \vu}\right) dt \approx \sum_{k=0}^{N-2} \frac{ \vk^2 \vkp^2}{\vt^6\D} \norms{\ukp - \uk}
\end{equation}

where $\vt = \frac{1}{\D}\ik \vhc(z(t)) dt$. Likewise, we have an approximation of the rate of change of the heading angle as a quadratic function using the substituted input variables. The constraints (\ref{eq:u_max}) becomes
\begin{equation}
\frac{\norms{ \ukp - \uk}}{\D\vt^2} \le \ps^2
\end{equation}
We discretize the dynamics without loss of precision using the state-transition matrix\cite{Reynolds2020AGuidance}. In the time-invariant case, we can recover its explicit formulation. With piece-wise affine input from \eqref{eq:input_piecewise_linear}, the dynamics constraint (\ref{eq:dyn}) can be expressed as
\begin{equation}
        \xkp = A_d \xk + B_k^- \uk+ B_k^+\ukp + \Wk
        \label{eq:lti}
\end{equation}
where $\Wk$ represents the integrated influence of the wind between $\tk$ and $\tkp$ and the state transition matrices are

\begin{equation}
    A_d =
    \begin{bmatrix}
    1&0\\0&1\\
    \end{bmatrix}
    \quad
    B^-_k = 
    \frac{\D}{2}
    \begin{bmatrix}
    1&0\\0&1\\
    \end{bmatrix}
    \quad
    B^+_k = 
    \frac{\D}{2}
    \begin{bmatrix}
    1&0\\0&1\\
    \end{bmatrix}
\end{equation}
\subsubsection{Linearization of constraints}
For the problem to be convex, the last step is to linearize the equality constraint (\ref{eq:nonL_cons}). Using a first-order truncated Taylor series expansion around a known previous iteration $\ul$ gives an approximated linear constraint
\begin{equation}
    \norm{\ul} - \vh + \frac{\ul\t}{\norm{\ul}}(\vu - \ul) = 0
    \label{eq:lin}
\end{equation}

Since this approximation is valid only when $\ul \approx \vu$, we have to impose $\norm{\uk - \bar{\v u}_k }\le\epsilon_u$ for each iteration and collocation point with a small enough $\epsilon_u$.

\subsection{Algorithm and convex programming}
Based on all the previous transformations, we can transform the Problem (\ref{eq:cvx_pa}) into a discrete convex problem solved for each iteration. At the iteration (n+1), we solve
    \begin{subequations}
 \begin{align}
\underset{\xk,\uk}{\text{ minimize }}& J = \boldsymbol\alpha\t 
    \begin{bmatrix}
    \norm{\v x_r(\tf) - \pf}\\
    \v u(\tf)\t \uf+1\\
    \end{bmatrix} + \sum_{k=0}^{N-2} \frac{ \vk^2 \vkp^2}{\vt^6\D} \norms{\ukp - \uk} &\textit{\quad cost function}\\
\text{subject to }&        
\xkp =   A_d \xk + B_k^- \uk+ B_k^+\ukp +\Wk&\textit{\quad dynamics}\label{eq:dyna_sys_eq}\\
&\norm{\unk} - \vk + \frac{\unk\t}{\norm{\unk}}(\uk - \unk) \ge -\epsilon_h & \textit{\quad substituted input}\label{eq:sub2}\\
&\norm{\uk} - \vk \le \epsilon_h&\textit{\quad substituted input}\label{eq:sub1}\\
&\norms{\frac{ \ukp - \uk}{\D}} \frac{1}{\vt^2} \le \ps^2 &\textit{\quad heading angle rate}\\ 
&     \norm \duk \le \epsilon_u&\textit{\quad linearization validity}\\
&    \duk = \uk - \unk&\textit{\quad definition}\\
&    \v x_0 = \xz & \textit{\quad initial condition}\\
&    \v u_0 = \uz&\textit{\quad initial condition}\label{eq:end}
\end{align}
\label{eq:eq1}
    \end{subequations}
    
The equality constraint (\ref{eq:lin}) have been replaced by two inequalities (\ref{eq:sub1}) and (\ref{eq:sub1}) with $\epsilon_h$ close to 0, to give the solver more flexibility to converge. The parameter $\epsilon_u$ constrain the gap between two consecutive iterations, trading off between the quality of the linearization and the convergence speed. The cost function $J$ is evaluated between each iteration and the convergence of the first SCP is assumed when its variation reaches a chosen threshold. At that stage, we switch to the second SCP of the algorithm where the parameter $\epsilon_h$ becomes a variable to be minimized and is added to the cost function. At the iteration (n+1), we have

    \begin{subequations}
                \begin{align}
\underset{\xk,\uk, \varepsilon_h}{\text{ minimize }} J& = \boldsymbol\alpha\t 
    \begin{bmatrix}
    \norm{\v x_r(\tf) - \pf}\\
    \v u(\tf)\t \uf+1\\
    \end{bmatrix}+ \sum_{k=0}^{N-2} \frac{ \vk^2 \vkp^2}{\vt^6\D} \norms{\ukp - \uk} + \alpha_5 \varepsilon_h &\textit{\quad cost function}\\
\text{subject to }  &      
(\ref{eq:dyna_sys_eq})&\textit{\quad dynamics}\\
&\norm{\unk} - \vk + \frac{\unk\t}{\norm{\unk}}(\uk - \unk) \ge -\varepsilon_h & \textit{\quad substituted input}\\
&\norm{\uk} - \vk \le \varepsilon_h&\textit{\quad substituted input}&\\ 
&(\ref{eq:sub1}) - (\ref{eq:end})&\textit{Left unchanged}
\end{align}
\label{eq:eq2}
    \end{subequations}
    
Overall, the method used stems from \cite{Mao2018AApplications}, where we have two Sequential Convex Programming, the first one giving an initial solution to the second, as proposed in \cite{Kelly2017AnCollocation}. The first SCP will yield a solution with a small dynamical infeasibility of $\epsilon_h$. The second SCP will include that value inside the cost function, so that the dynamical infeasibility is minimized. Implementing the second SCP is not essential and is only meant to improve the solution slightly. In practice, the selection of $\epsilon_h$ is driven by the following trade-off. When $\epsilon_h$ is small, the convergence of the SCP is slower but the resulting solution is more accurate. Oppositely, when $\epsilon_h$ is larger, less iterations will be needed and the resulting solution will be more dynamically accurate. In practice, the second SCP converges usually after a single iteration. In order to accelerate the convergence, we can also infer an initial trajectory based on a previous computed optimal trajectory, where the initial state is updated. The algorithm is summarised below. 

\begin{figure}[ht!]
     \centering
     \includegraphics[width = 0.8\textwidth]{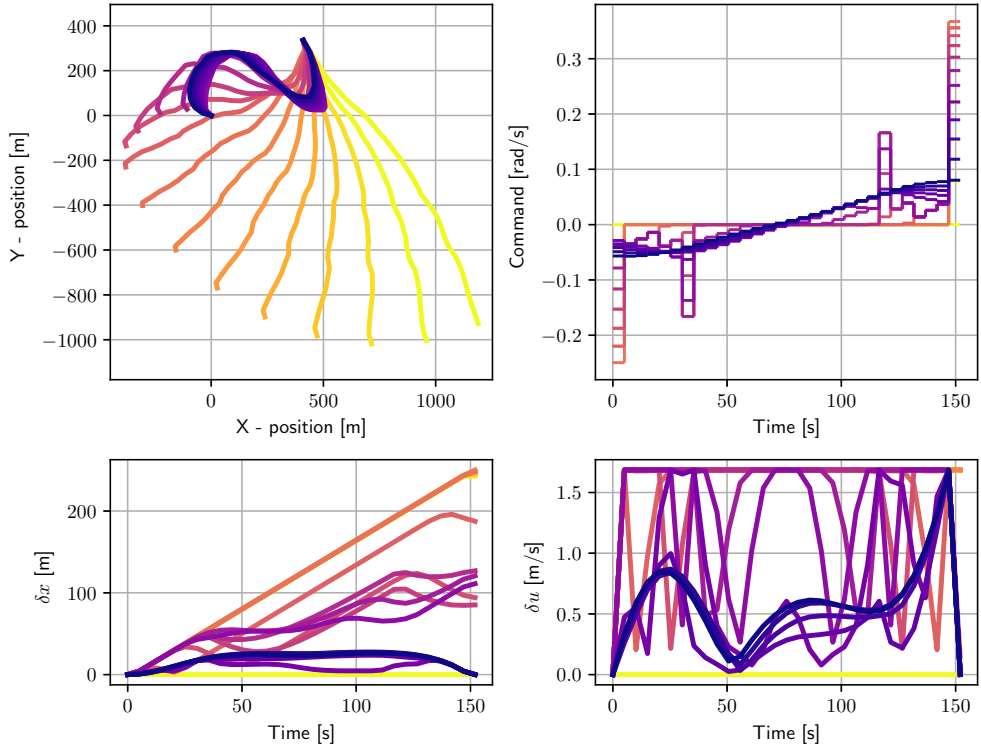}
     \caption{Iterations required to convergence to a local minimum actuation energy from light yellow to dark blue. Each iteration is dynamically feasible.}
     \label{fig:convergence}
 \end{figure}

\begin{figure}[ht!]
     \centering
     \includegraphics[width = 0.8\textwidth]{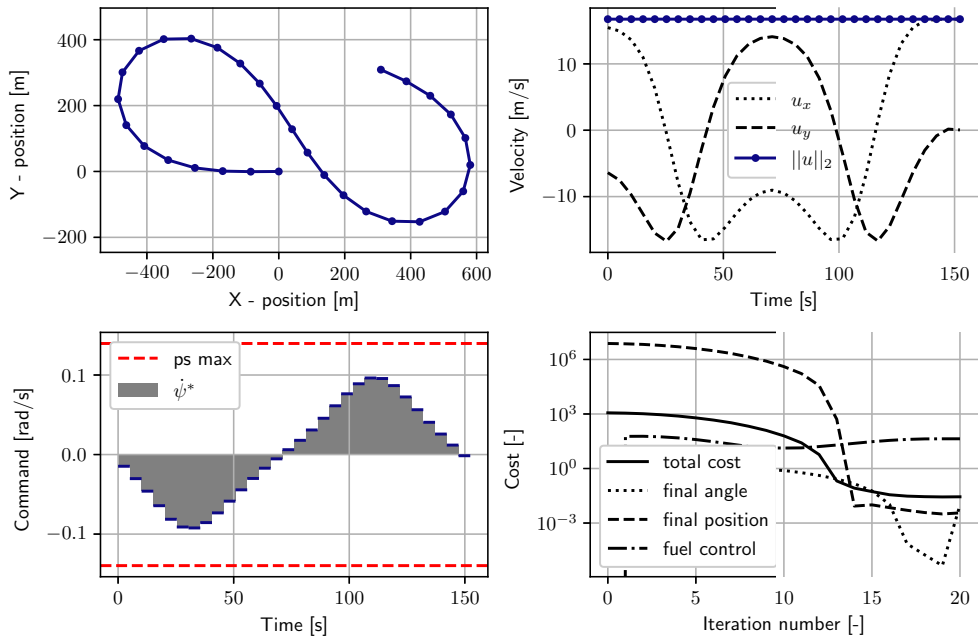}
     \caption{Illustration of the solution of the iterative algorithm, along with the convergence of the cost function}
     \label{fig:convergence_part2}
 \end{figure}

\begin{algorithm}[H]
\SetAlgoLined
\% Offline initialisation\;
Initialisation of $\epsilon_h$, $\epsilon_u$, $\boldsymbol \alpha$, $c_h$, $c_\rho$, $c_e$ \% mission independent\;
Initialisation of $\ps$, $\pf$, $\psi^f$;\% mission dependent\;
\% Online initialisation\;
Initialisation of $\vk$, $\vt$, $W_k$, $\uz$, $\xz$, $\tf$, $\tz$, $\D$, $A_d$, $B_k^-$, $B_k^+$ based on the navigation sensors\;
Initial guess of the trajectory $\un$. n=0.\;
 \While{J has not converged}{
  Linearization of the problem around $\un$\;
  Resolution of the problem (\ref{eq:eq1})\;
  n=n+1\;
  Update $\un$ with the new solution\;
 }
 Output $\unp$ as initial guess\;
 \While{J has not converged}{
 Linearisation of the problem around $\un$\;
 Resolution of the problem (\ref{eq:eq2})\;
 n=n+1\;
 Update $\un$ with the new solution\;
 }
 Output $\v u_r^*(t) = \unp$ and $\v x_r^*(t)$ as the solution of the Problem (\ref{eq:cvx_pa})\;
 Evaluate Eq.(\ref{eq:conv_states}) to recover the solution of Problem (\ref{eq:opt_problem})\;
 \KwResult{Feasible reference trajectory for the guidance problem}
 \caption{Successive convexification for the parafoil guidance problem}
\end{algorithm}
To start the iterative process, we choose a constant initial guess for $\vu$, propagating from the initial state, such as $\bar{\v u}^{(0)}(t) = \uz \quad  \forall t \in [\tz,\tf]$ and we derive $\xk$ accordingly with Eq.(\ref{eq:lti}). Each iteration, including the initial guess will yield a feasible trajectory, oppositely to \cite{Reynolds2020AGuidance}, where an artificial input is included to deal with the artificial infeasabilities. The method presented here avoid using the artificial input from \cite{Reynolds2020AGuidance}, by tailoring the sCVX algorithm to the more specific parafoil landing problem. To reduce the number of iterations required to converge, further development could focus on a better feasible initial guess. On Fig.\ref{fig:convergence}, we see that each iteration get the resulting trajectory closer to the targeted final state in $\pf = [0,0]^T$ with the heading angle $\psi^f$ aligned with the X-axis. The variation of state $\delta x$ and input $\delta u$ and also converges. The shape of the resulting trajectory is not obvious since it takes into account the wind perturbation. Because of the choice of discretization, the asymmetrical deflection to be applied on the parafoil $u^*(t)$ is piecewise-constant. The Fig.\ref{fig:convergence_part2} shows an optimal trajectory with the associated constraints and converging total cost.
 
\subsection{Controller}
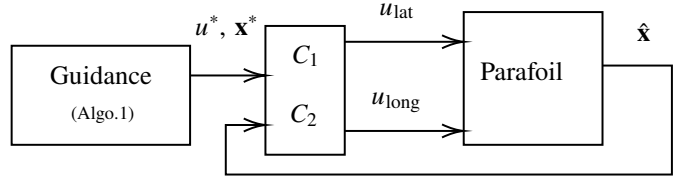
\begin{figure}[ht!]
    \centering
    \tikzset{every picture/.style={line width=0.75pt}} 

\begin{tikzpicture}[x=0.75pt,y=0.75pt,yscale=-1,xscale=1]

\draw   (133,20) -- (173,20) -- (173,85) -- (133,85) -- cycle ;
\draw   (5,30) -- (95,30) -- (95,80) -- (5,80) -- cycle ;
\draw   (233,13) -- (303,13) -- (303,80) -- (233,80) -- cycle ;
\draw    (303,40) -- (338,40) -- (338,95) -- (113,95) -- (113,70) -- (131,70) ;
\draw [shift={(133,70)}, rotate = 180] [color={rgb, 255:red, 0; green, 0; blue, 0 }  ][line width=0.75]    (10.93,-3.29) .. controls (6.95,-1.4) and (3.31,-0.3) .. (0,0) .. controls (3.31,0.3) and (6.95,1.4) .. (10.93,3.29)   ;
\draw    (173,28) -- (231,28) ;
\draw [shift={(233,28)}, rotate = 180] [color={rgb, 255:red, 0; green, 0; blue, 0 }  ][line width=0.75]    (10.93,-3.29) .. controls (6.95,-1.4) and (3.31,-0.3) .. (0,0) .. controls (3.31,0.3) and (6.95,1.4) .. (10.93,3.29)   ;
\draw    (173,73) -- (231,73) ;
\draw [shift={(233,73)}, rotate = 180] [color={rgb, 255:red, 0; green, 0; blue, 0 }  ][line width=0.75]    (10.93,-3.29) .. controls (6.95,-1.4) and (3.31,-0.3) .. (0,0) .. controls (3.31,0.3) and (6.95,1.4) .. (10.93,3.29)   ;
\draw    (95,45) -- (131,45) ;
\draw [shift={(133,45)}, rotate = 180] [color={rgb, 255:red, 0; green, 0; blue, 0 }  ][line width=0.75]    (10.93,-3.29) .. controls (6.95,-1.4) and (3.31,-0.3) .. (0,0) .. controls (3.31,0.3) and (6.95,1.4) .. (10.93,3.29)   ;

\draw (153.5,35) node   [align=left] {$\displaystyle C_{1}$};
\draw (50,45) node   [align=left] {Guidance};
\draw (152.5,65) node   [align=left] {$\displaystyle C_{2}$};
\draw (240,37) node [anchor=north west][inner sep=0.75pt]   [align=left] {Parafoil};
\draw (319,17) node [anchor=north west][inner sep=0.75pt]   [align=left] {$\displaystyle \hat{\mathbf{x}}$};
\draw (189,7) node [anchor=north west][inner sep=0.75pt]   [align=left] {$\displaystyle u_{\text{lat}}$};
\draw (185,52) node [anchor=north west][inner sep=0.75pt]   [align=left] {$\displaystyle u_{\text{long}}$};
\draw (96,13) node [anchor=north west][inner sep=0.75pt]   [align=left] {$\displaystyle u^{*} ,$};
\draw (116,13) node [anchor=north west][inner sep=0.75pt]   [align=left] {$\displaystyle \mathbf{x}^{*}$};
\draw (51,65) node   [align=left] {{\scriptsize (Algo.1)}};

\end{tikzpicture}
    \caption{Architecture of the guidance and control algorithm}
    \label{fig:control}
\end{figure}

The controller is needed to compensate for the modeling errors and approximations and unmodelled perturbations and noise. The algorithm above gives a reference for the input $u^*(t)$ and for the states $\v x^*(t)$, used as a feedforward to the controller illustrated on the Fig.\ref{fig:control}. It operates independently on both the lateral and longitudinal channel to ensure that the parafoil has respectively a minimum lateral and longitudinal error. The symmetrical and asymmetrical deflection of the parafoil controls the rate of fall and the rate of turn of the parafoil. The longitudinal control compensates for the error in the estimation of the time of flight and rate of fall. We assume that the parafoil navigation provide $\hat{\v x}$. Inspired by \cite{Rademacher2009}, for the lateral control, we have the following asymmetrical deflection

\begin{equation}
    u_\text{lat}(t) = u^*(t) + C_1(\hat{\v x}(t), {\v x^*(t)})
\end{equation}

and for the longitudinal control, we have the symmetrical deflection

\begin{equation}
    u_\text{long}(t) = C_2(\hat{\v x}(t), {\v x^*(t)})
\end{equation}

The control algorithm is implemented only for illustrative purposes and further development could focus on an optimally tuned controller. $C_1$ and $C_2$ are controller respectively on the cross-track error and the along-track error.
\section{Results}
\label{sec:results}
To assess the performances of the algorithm, a 6-DOF model developed by Rademacher\cite{Rademacher2009} was used as a benchmark of the physical reality, including the saturations on the actuators. The 4-DOF model used in Eq.(\ref{eq:4dof}) was developed independently.

    \subsection{Monte-Carlo} 
\begin{figure}[ht!]
        \centering
        \includegraphics{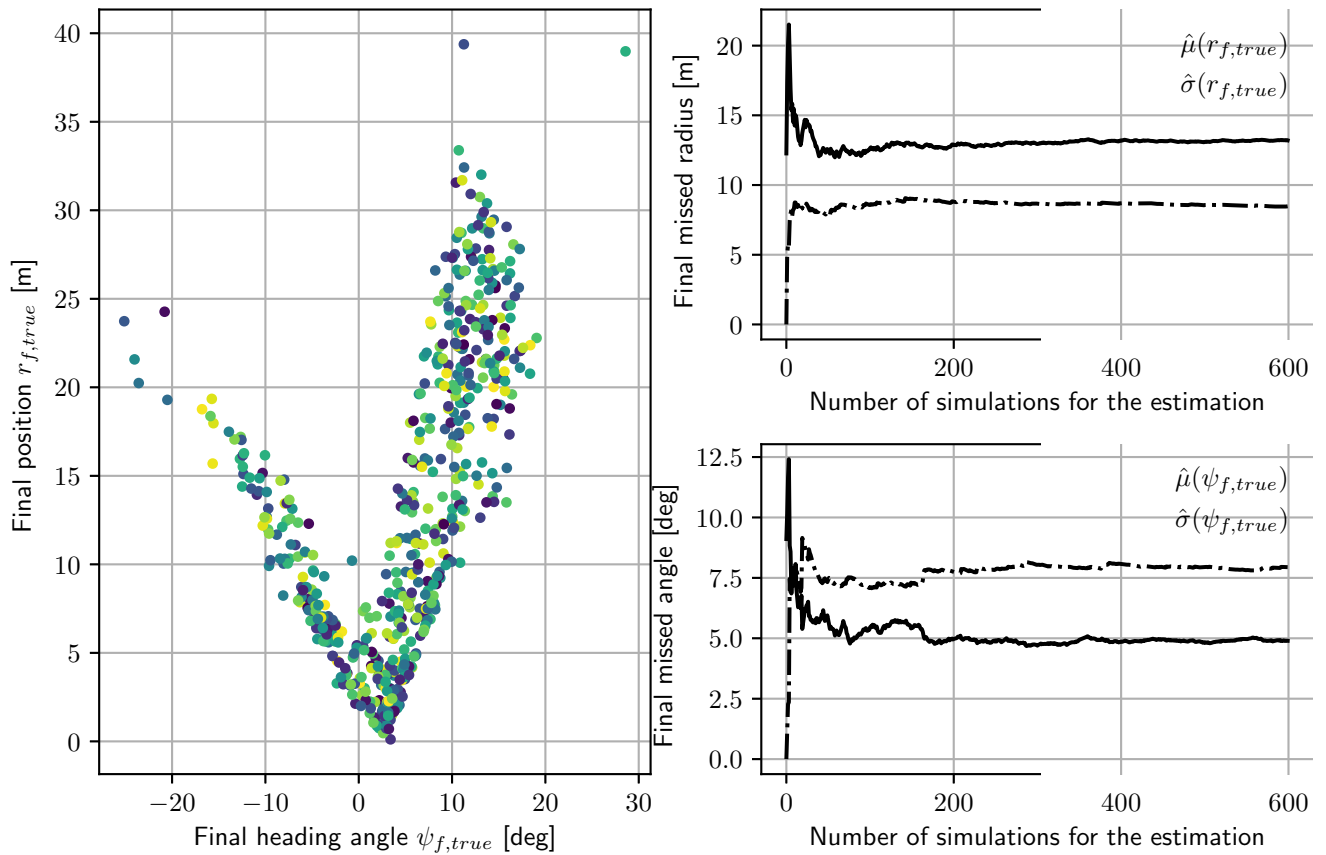}
        \caption{Final position and heading angle with open-loop guidance, lateral and longitudinal control, perfect knowledge of the strong wind (left). Estimation of the mean and variance of the missed distance and missed heading angle (right).}
        \label{fig:landing}
    \end{figure}

\begin{figure}
     \centering
     \includegraphics{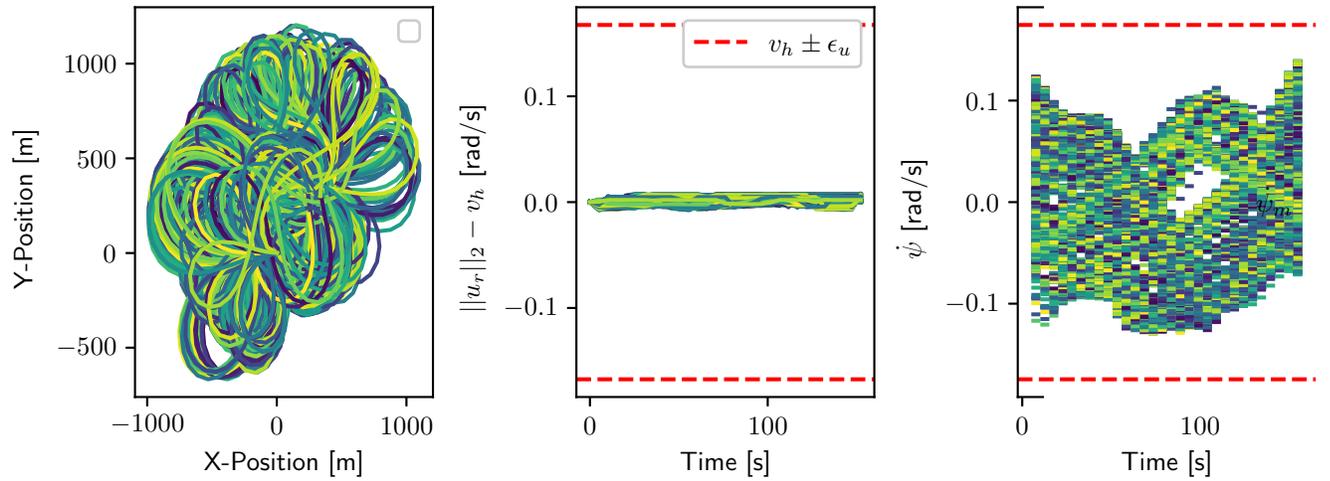}
     \caption{Monte-Carlo simulation with a wide range of initial conditions $\xz$, where we can see that the constraints are respected.}
     \label{fig:solution}
 \end{figure}

The 600 runs of the Monte-Carlo simulation are used to show the good convergence properties of the algorithm exposed for a large set of initial conditions. On the Fig.\ref{fig:landing}, we can see that all trajectories have led the parafoil close to the landing site. In this case, a single reference trajectory have been computed and tracked with the aforementioned controller. In practice, better results can be achieved when a new trajectory is computed when the parafoil is too far from the reference. Additionally, we can see on the right hand side that the algorithm is consistent. The mean missed distance and mean missed heading angle reaches a steady value after only a few simulations, indicating that the algorithm does not produce outliers. On the Fig. \ref{fig:solution}, we can see that regardless of the initial conditions in heading angle and position, a trajectory satisfying the final condition is found in less than 35 iterations. We can also point out that each iteration takes on average \SI{0.004}{\second} to be computed on a regular laptop, i7 16Go RAM. 

\subsection{Comparison}
\begin{figure}[ht!]
        \centering
        \includegraphics[width = 0.6\textwidth]{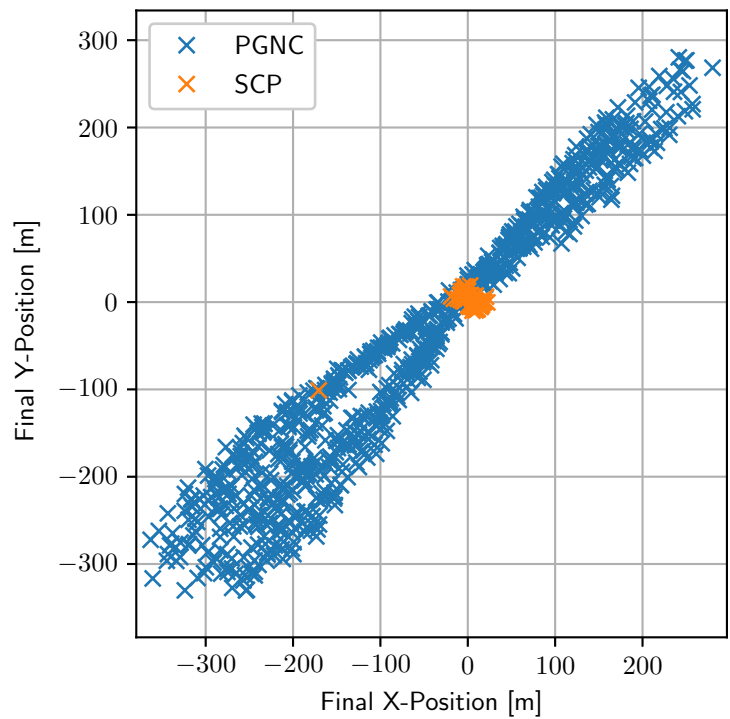}
        \caption{Comparison of the missed landing distance between the traditional offline algorithm 'PGNC' and the online optimization 'SCP' presented in this article, with known wind distribution}
        \label{fig:finalposcomp}
    \end{figure}

The online trajectory generation methods is compared against a more traditional method; the energy management. The energy management method is tested for a 4-DOF X-38 parafoil model calibrated with flight data while the aforementioned algorithm is used on a comparable 6-DOF parafoil model, with the same wind model in both cases. With similar initial conditions, we can observe the difference in landing precision on the Fig. \ref{fig:finalposcomp}. The final heading angle constraint has been removed for higher consistency and comparability, since the EM method does not target a specific heading angle. The dispersion of the landing precision is much smaller for the online optimization method.

\section{Conclusion}
The parafoil guidance problem have been solved by the application of a sequential convex programming. Using a substitution for the trigonometric function and linearization, it is possible to convert the non-convex problem into a sequence of convex problems, suitable for onboard implementation and real-time use. Especially, each iteration yields a feasible solution of the physical problem in polynomial time and can be used as a sub-optimal feedforward trajectory, to comply with real-time constraints. The algorithm takes into account the influence of the altitude-varying density of the air on the aerodynamics and requires only the control saturation level for the parafoil model. It also considers the wind distribution and find a locally optimal solution, minimizing the control energy or the missed landing or both when possible. The algorithm is tested on a 6-DOF parafoil model. The navigation error is not included in the simulation. Further test will include hardware-in-the-loop simulation.

\section*{Acknowledgments}
The authors thank all the TEC-SA division within the European Space Agency for their continuous support, as well as the the numerous comments and feedback of Dr. Carlo Alberto Pascucci from Embotech and the precious help from Hans Strauch. This work has been carried out during a Young Graduate Trainee Program at the European Space Agency. 

\newpage
\newpage
\bibliography{bibli}

\end{document}